\documentstyle{elsart}
\begin{document}
\begin{frontmatter}
\title{Algorithmic Integrability Tests for Nonlinear Differential
and Lattice Equations\thanksref{NSF}}

\author{Willy Hereman\thanksref{email1}\thanksref{email2}},
\author{\"{U}nal G\"{o}kta\c{s}\thanksref{email1}\thanksref{wolfram}}, 
\author{Michael D. Colagrosso\thanksref{email1}}

\address{
Department of Mathematical and Computer Sciences,
Colorado School of Mines, Golden, CO 80401-1887, U.S.A.} 

\author{and Antonio J. Miller\thanksref{email3}}

\address{
Advanced Sensors and Control Department, The Applied Research Laboratory, 
The Pennsylvania State University, State College, PA 16804-0030, U.S.A.}

\thanks[NSF]{Research supported in part by NSF
under Grant CCR-9625421.}
\thanks[email1]{E-mail: $\{$whereman,ugoktas,mcolagro$\}$@mines.edu}
\thanks[email2]{Corresponding author.
Phone: (303) 273 3881; Fax: (303) 273 3875.}
\thanks[wolfram]{Supported in part by Wolfram Research, Inc.}
\thanks[email3]{E-mail: ajmiller@sabine.acs.psu.edu}

\begin{abstract}
Three symbolic algorithms for testing the integrability of polynomial
systems of partial differential and differential-difference 
equations are presented. 
The first algorithm is the well-known Painlev\'e test, which is applicable
to polynomial systems of ordinary and partial differential equations. 
The second and third algorithms allow one to explicitly compute polynomial 
conserved densities and higher-order symmetries of nonlinear evolution 
and lattice equations. 

The first algorithm is implemented in the symbolic syntax of both
{\it Macsyma} and {\it Mathematica}. 
The second and third algorithms are available in {\it Mathematica}. 
The codes can be used for computer-aided integrability testing of nonlinear
differential and lattice equations as they occur in various branches
of the sciences and engineering. 
Applied to systems with parameters, the codes can determine the 
conditions on the parameters so that the systems pass the Painlev\'e 
test, or admit a sequence of conserved densities or higher-order symmetries.
\end{abstract}

\begin{keyword}
Integrability; Painlev\'e test; Conservation law; Invariant; Symmetry; 
Differential equation; Lattice
\end{keyword}
\end{frontmatter}

\section{Introduction}
During the last three decades, the study of integrability, invariants, 
symmetries, and exact solutions of nonlinear ordinary and partial
differential equations (ODEs, PDEs) and differential-difference 
equations (DDEs) has been the topic of major research projects in the 
dynamical systems and the soliton communities. 

Various techniques have been developed to determine whether or not
PDEs belong to the privileged class of completely integrable equations
\cite{flanewtab}. 
One of the most successful and widely applied techniques is the 
Painlev\'e test, named after the French mathematician 
Paul Painlev\'e (1863-1933) \cite{levwin}, who classified 
second-order differential equations that are globally integrable in terms 
of elementary functions by quadratures or by linearization.  
In essence, the Painlev\'e test verifies whether or not solutions of
differential equations in the complex plane are single-valued in the 
neighborhood of all their movable singularities. 

To a large extent, the Painlev\'e test is algorithmic, yet very cumbersome 
when done by hand. In particular, the verification of the compatibility 
conditions is assessed by many practitioners of the Painlev\'e test 
as a painstaking computation. 
In addition to the tedious verification of self-consistency 
(or compatibility) conditions, computer programs are helpful at 
exploring {\it all} possibilities of balancing singular terms. 
Indeed, the omission of one or more choices of ``dominant behavior" can 
lead to wrong conclusions \cite{congraram,ramgrabou}.
% \cite{cla1,cla2,cla3,ramgrabou,ramgra}.

We therefore developed the programs {\it painsing.max} and 
{\it painsys.max} \cite{oursite} 
(both in {\it Macsyma} syntax \cite{macsyma}), and {\it painsing.m} 
and {\it painsys.m} in {\it Mathematica} language \cite{SW}, 
that perform the Painlev\'e test for polynomial systems of ODEs and PDEs.
In this paper we demonstrate the above codes by analyzing a few prototypical 
nonlinear equations and systems, such as the Boussinesq and nonlinear 
Schr\"odinger equations, a class of fifth-order KdV equations, and the 
Hirota-Satsuma and Lorenz systems.

Our computer code does not deal with the theoretical shortcomings of 
the Painlev\'e test as identified by Kruskal and others 
\cite{kru,krucla,krujoshal}. 
Thus far, we have implemented the traditional Painlev\'e test
\cite{herang,herzhu}, and not yet incorporated the latest 
advances in Painlev\'e type methods, such as the poly-Painlev\'e test 
\cite{krucla} or other generalizations \cite{kru,krujoshal}.  
Neither did we code the weak Painlev\'e test \cite{ramdorgra,ramgrabou}
or other variants \cite{conte,contebook,conforpic,gor}.
Furthermore, we do not have code for the singularity confinement method
\cite{grarampap}, 
i.e. an adaptation of the Painlev\'e test that allows one to test the 
integrability of difference equations. 

Among the various alternatives to establish the integrability 
\cite{flanewtab} of nonlinear PDEs and DDEs, the search for 
conserved densities and higher-order symmetries is particularly appealing. 
Indeed, in this paper we will give algorithms that apply to both 
the continuous and semi-discrete cases. We implemented these algorithms
\cite{jsc,physicad,baltzer} in {\it Mathematica}, 
but they are fairly simple to code in other computer algebra languages
(see \cite{jsc,baltzer}). 

Our algorithms are based on the concept of dilation (scaling) invariance.
That inherently limits their scope to polynomial conserved 
densities and higher-order symmetries of polynomial systems.  
Although the existence of a sequence of conserved densities 
predicts integrability, the nonexistence of polynomial conserved 
quantities does not preclude integrability.
Indeed, integrable PDEs or DDEs could be disguised with a coordinate 
transformation in DDEs that no longer admit conserved densities of 
polynomial type \cite{ASandRY}. The same care should be taken in 
drawing conclusions about non-integrability based on the lack of 
higher-order symmetries, or equations failing the Painlev\'e test. 

Apart from integrability testing, the knowledge of the explicit form
of conserved densities and higher-order symmetries is useful.
For instance, with higher-order symmetries of integrable systems, 
one can build new completely integrable systems, or discover connections
between integrable equations and their group theoretic origin. 

Explicit forms of conserved densities are useful in the numerical solution 
of PDEs or DDEs. 
In solving DDEs, which may arise from integrable discretizations of PDEs, 
one should check that conserved quantities indeed remain constant.
In particular, the conservation of a positive definite quadratic quantity 
may prevent nonlinear instabilities in the numerical scheme. 

Our integrability package {\it InvariantsSymmetries.m} \cite{invsymmsoft}
in {\it Mathematica\/} automates the tedious computation of closed-form 
expressions 
for conserved densities and higher-order symmetries for both PDEs and DDEs.
Applied to systems with parameters, the package determines the conditions 
on these parameters so that a sequence of conserved densities or 
symmetries exists. 
The software can thus be used to test the integrability of classes of 
equations that model various wave phenomena. 
Our examples include a vector modified KdV equation, the extended 
Lotka-Volterra and relativitic Toda lattices, and the Heisenberg
spin model. 

The conserved densities and symmetries presented in this paper were 
obtained with {\it InvariantsSymmetries.m}.
\vfill
\section{Symbolic Program for the Painlev\'e Test}
\subsection{Purpose}
We focus on PDEs. As originally formulated by Ablowitz {\it et al.} 
\cite{ablramseg1,ablramseg2}, the Painlev\'e conjecture asserts 
that all similarity reductions of a completely integrable PDE 
should be of Painlev\'e-type; i.e.\ its solutions should have 
no movable singularities other than ``poles" in the complex plane. 

A later version of the Painlev\'e test due to Weiss {\it et al.} 
\cite{weitabcar} allows testing of the PDE directly, without recourse to 
the reduction(s) to an ODE.
A PDE is said to have the Painlev\'e property \cite{ablcla} if its 
solutions in the complex plane are single-valued in the neighborhood of 
all its movable singularities. In other words, the equation must have
a solution without any branching around the singular points whose
positions depend on the initial conditions.
For ODEs, it suffices to show that the general solution has no worse 
singularities than movable poles, or that no branching occurs around 
movable essential singularities.  

A three step-algorithm, known as the Painlev\'e test, allows one to verify
whether or not a given nonlinear system of ODEs or PDEs with (real) 
polynomial terms fulfills the necessary conditions for having the 
Painlev\'e property.
Such equations are prime candidates for being completely integrable.

There is a vast amount of literature about the test and its applications 
to specific ODEs and PDEs. Several well-documented surveys 
\cite{cartab,conte,flanewtab,krucla,krujoshal,laksah,muscon1,newtabzen}
and books \cite{clarksonbook,contebook,steeul} discuss subtleties 
and pathological cases of the test that are far beyond the scope of this 
article. Other survey papers \cite{flanewtab,gibetal,muscon2} 
deal with the many interesting by-products of the Painlev\'e test.
For example, they show how a truncated Laurent series expansion of the type 
introduced below, allows one to construct Lax pairs, B\"acklund and Darboux
transformations, and closed-form particular solutions of PDEs. 

\subsection{Algorithm for a Single Equation}
We briefly outline the three steps of the Painlev\'e test for a single PDE,
\begin{equation}\label{generalpde}
{\cal F} (x,t,u(x,t)) = 0,
\end{equation} 
in two independent variables $x$ and $t$. 
Our software can handle the four independent variables $(x,y,z,t).$
Throughout the paper we will use the notations 
\begin{equation} 
\label{notations}
u_t = \frac{\partial u}{\partial t}, \quad
u_{nx} = \frac{\partial^n u}{\partial x^n}, \quad
u_{tx} = \frac{\partial^2 u}{\partial t \partial x}, \quad {\rm etc.}
\end{equation}
In the approach proposed by Weiss, the solution $u(x,t)$, expressed as 
a Laurent series 
\begin{equation} 
\label{laurent}
u(x,t) = g^{\alpha}(x,t) \sum_{k=0}^{\infty} u_{k}(x,t) \; g^{k}(x,t), 
\end{equation} 
should be single-valued in the neighborhood of a non-characteristic, 
movable singular manifold $g(x,t), $ which can be viewed as the 
surface of the movable poles in the complex plane.
In (\ref{laurent}), $u_{0} (x,t) \ne 0$, $\alpha$ is a negative integer,
and $u_{k}(x,t) $ are analytic functions in a neighborhood of $g(x,t).$

Note that for ODEs the singular manifold is $g(x,t) = x - x_0,$ 
where $x_0$ is the initial value for $x$.
For PDEs, if $u(x,t)$ has simple zeros and $g_{x}(x,t) \ne 0,$ 
one may apply the implicit function theorem near the singularity manifold 
and set $g(x,t) = x - h(t),$ for an arbitrary function $h(t)$ 
\cite{kruramgra,ramgrabou}. 
This so-called the Kruskal simplification, considerably reduces the length 
of the calculations.
% \vfill
% \newpage
\vskip 1pt 
\noindent
The Painlev\'e test proceeds in three steps:
\vskip 1pt
\noindent
{\bf Step 1: Determine the dominant behavior}
\vskip 1pt
\noindent
Determine the negative integer $\alpha$ and $u_0$ from the 
leading order ``ansatz". This is done by balancing the minimal power 
terms after substitution of $u \propto u_0 g^{\alpha}$ into the given PDE.
There may be several branches for $u_0,$ and for each the next two 
steps must be performed. 
\vskip 1pt
\noindent
{\bf Step 2: Determine the resonances}
\vskip 1pt
\noindent
For a selected $\alpha$ and $u_0,$ calculate the non-negative integer 
powers $r$, called the {\it resonances}, at which arbitrary functions 
$u_r$ enter the expansion. 
This is done by requiring that $u_r$ is arbitrary after substitution
of $u \propto u_0 g^{\alpha} + u_r g^{\alpha+r}$ into the equation, 
only retaining its most singular terms.
The coefficient $u_r$ will be arbitrary if its coefficient equals zero. 
The integer roots of the resulting polynomial must be computed.
The number of roots, including $r=-1,$ should match the order of the 
given PDE. The root $r=-1$ corresponds to the arbitrariness of the 
manifold $g(x,t)$,
\vskip 1pt
\noindent
{\bf Step 3: Verify the correct number of free coefficients}
\vskip 1pt
\noindent
Verify that the correct number of arbitrary functions $u_r$ indeed 
exists by substituting the truncated expansion 
\begin{equation}
\label{fulllaurent}
u(x,t) = g^{\alpha} {\sum_{k=0}^{{\rm rmax}} u_{k} (x,t) g^{k} (x,t)} 
\end{equation}
into the PDE, where $rmax$ is the largest resonance.
At non-resonance levels, determine all $u_k.$
At resonance levels, $u_r$ should be arbitrary, and since we are 
dealing with a nonlinear equation, a {\it compatibility condition} must 
be verified. 

An equation for which the above steps can be carried out consistently
and unambiguously, is said to have the Painlev\'e property and is 
conjectured to be completely integrable. 
This entails that the solution has the necessary number of free coefficients 
$u_r,$ and that the compatibility condition at each of these resonances 
is unconditionally satisfied. 

The reader should be warned that the above algorithm does not detect 
essential singularities and therefore cannot determine whether or not
branching occurs about these. 
So, for an equation to be integrable it is {\it not} {\it sufficient} 
that it passes the Painlev\'e test. 
Neither it is {\it necessary}. Indeed, there are integrable equations, 
such as the Dym-Kruskal equation, $u_t = u^3 u_{3x},$ that do not pass the 
Painlev\'e test, yet, by a complicated change of variables can 
be transformed into an integrable equation.  

\subsection{Algorithm for Systems}
The generalization of the algorithm to systems of ODEs and PDEs is obvious.
Yet, it is non-trivial to implement. One of the reasons is that the 
major symbolic packages do not handle inequalities well.

With respect to systems, our code is based on the above three 
step-algorithm but generalized to systems, as it can be found 
in \cite{kruramgra,ramgrabou,steeul}. 
In these papers there is an abundance of worked examples that served as 
test cases.

For example, given a system of first-order ODEs, 
\begin{equation}
\label{odesystem}
\frac{d u_i }{d x} = G_i ( u_1 , u_2 , ... , u_n ; x ) ,
\quad\quad i = 1,2, ..., n,
\end{equation}
one introduces a Laurent series for every dependent variable $u_i(x):$ 
\begin{equation}
\label{odeseries}
u_i = (x - x_0 )^{\alpha_i} \sum_{k=0}^{\infty} u_k^{(i)} (x - x_0)^k.
\end{equation}
The computer program must carefully determine all branches of dominant 
behavior corresponding to various choices of $\alpha_i$ and/or $u_0^{(i)}.$ 
For each branch, the single-valuedness of the corresponding Laurent 
expansion must be tested (i.e. the resonances must be computed and the 
compatibility conditions must be verified). 
All the details can be found in \cite{flanewtab,kruramgra,ramgrabou}.

Singularity analysis for PDEs is nontrivial \cite{krujoshal} and 
the Painlev\'e test should be applied with extreme care. 
Notwithstanding, our software automatically performs the 
{\it formal} steps of the Painlev\'e test for systems of ODEs and PDEs. 
The examples in section 3.2 illustrate how the code works. 
Careful analysis of the output and drawing conclusions about integrability 
should be done by humans. 
Some subtleties of the mathematics of the Painlev\'e test of systems of 
PDEs were also dealt with in \cite{chuchutab,doksak,hirgraram,tamsah}.

\section{Examples of the Painlev\'e Test}
\subsection{Single Equations}

Numerous examples of the Painlev\'e test for ODEs can be 
found in the review papers. We turn our attention to PDEs. 
Using our software package {\it painsing.max} or {\it painsing.m} 
one can determine the conditions under which the equation
\begin{equation}
\label{cylindricalkdv}
u_{tx} + a(t) u_x + 6 u u_{2x} + 6 u_x^2 + u_{4x} = 0, 
\end{equation}
passes the Painlev\'e test.

For (\ref{cylindricalkdv}), $\alpha = -2$ and $u_0 = - 2 g_x^2.$
Apart from $r=-1,$ the roots are $r=4, 5,$ and $6.$
The latter three are resonances. Furthermore, 
\begin{eqnarray}
\label{coefcylkdv}
u_1 &=& 2 g_{2x}, \quad \;
u_2 = -\frac{1}{6 g_x^2} (4 g_x g_{3x} -3 g_{2x}^2 + g_t g_x), \\ 
u_3 &=& \frac{1}{6 g_x^4}( a(t) g_x^3 + g_x^2 g_{4x} -4 g_x g_{2x} g_{3x} 
        + 3 g_{2x}^2 -  g_t g_x g_{2x} + g_{tx} g_x^2), 
\end{eqnarray}
and $u_4$ and $u_5$ are indeed arbitrary since the compatibility 
conditions at resonances $r=4$ and $r=5$ are satisfied identically. 

The compatibility condition at resonance $r=6$ is
% \begin{equation}
% \label{compatlevel6}
$ a_t + 2 a^2 = 0. $
% \end{equation}
Ignoring the trivial solution, we get $a = \frac{1}{2(t-t_0)}.$ 
Without loss of generality, we set $t_0=0$ and equation 
(\ref{cylindricalkdv}) becomes the {\it cylindrical} KdV equation 
which is indeed completely integrable \cite{ablcla,caldeg}.
Painlev\'e based investigations for integrable PDEs with space and 
time dependent coefficients are given in 
\cite{ablcla,clarksonpap,herproc,herzhu}. 

Our Painlev\'e programs cannot automatically test a class of equations 
such as
\begin{equation} 
\label{classfifthkdv}
u_t + a u^2 u_x + b u_x u_{2x} + c u u_{3x} + u_{5x} = 0,
\end{equation}
with arbitrary (non-zero and real) parameters $a, b $ and $ c.$
The parameters affect the lowest coefficient in the Laurent expansion
in such as way that the roots $(r)$ cannot be computed, and the 
integrability conditions can no longer be tested. 

In (\ref{classfifthkdv}) there are four cases that are of particular interest:
\begin{itemize}
\item[] (i)   $a = \frac{3}{10}c^2$ and $b=2c \;\;\;\!$ (Lax equation), 
\vskip 4pt 
\item[] (ii)  $a = \frac{1}{5}c^2$ and $b=c \;\;\;\;\,$ (Sawada-Kotera equation), 
\vskip 4pt 
\item[] (iii) $a = \frac{1}{5}c^2$ and $b=\frac{5}{2}c \;\;\;\!\!$  
(Kaup-Kupershmidt equation), and 
\vskip 4pt 
\item[] (iv)  $a = \frac{2}{9}c^2$ and $b=2c \;\;\;\!$ (Ito equation). 
\end{itemize}
In Table 1 we list the results of the Painlev\'e test applied to these cases.
For the first three equations the compatibility conditions are satisfied
at all the resonances. These equations pass the test. 
For the Ito equation the compatibility conditions
are only satisfied at some of the resonances. The Ito equation fails the test.
The first three equations are known to be completely integrable. 
Ito's equation is not completely integrable.

The two other algorithms presented in this paper can determine the 
conditions (i), (ii) and (iii) that assure the complete integrability 
of (\ref{classfifthkdv}). 
The conserved densities and higher-order symmetries of 
(\ref{classfifthkdv}) can be found in \cite{jsc} and \cite{baltzer}. 
\vskip 0.01pt
\noindent
\begin{center}
% Start Table 1 Painlev\'e Analysis
% \begin{table}[hp] \caption{{\rm Conserved Densities 
% for the Heisenberg Spin Model}}
% \vskip .001pt
% \noindent
% \begin{footnotesize}
\begin{tabular}{ | l | l | l | l|} \hline 
Lax & Sawada-Kotera &$\!$Kaup$\!-\!$Kupershmidt$\!$&Ito  \\  [0.5ex] 
\hline
$\!\!(a,b,c)\!\!=\!\!(30,20,10)\!$ &$\!(a,b,c)\!=\!(5,5,5)\!$
& $\!(a,b,c)\!=\!(20,25,10)\!$ &$\!(a,b,c)\!=\!(2,6,3)\!$ \\
\hline
$\alpha = -2 $& 
$\alpha = -2 $&
$\alpha = -2 $& 
$\alpha = -2 $ \\
\hline
Branch 1 & Branch 1 & Branch 1 & Branch 1 \\
$u_0=-2 g_x^2$ &
$u_0=-6 g_x^2$ &
$u_0=-\frac{3}{2} g_x^2$ &
$u_0=-6 g_x^2$ \\
$\!r\!\!=\!\!-1,2,5,6,8\!$ & 
$\!r\!\!=\!\!-1,2,3,6,10\!$ & 
$\!r\!\!=\!\!-1,3,5,6,7\!$ & 
$\!r\!\!=\!\!-1,3,4,6,8\!$ \\
OK for $r \ge 2$ &
OK for $r \ge 2$ &
OK for $r \ge 3$ &
OK at $r\!=\!3$ \\
& & &Not at $\!r\!=\!4,6,8$ \\
\hline
Branch 2 & Branch 2 & Branch 2 & Branch 2 \\
$u_0=-6 g_x^2$  &
$u_0=-12 g_x^2$ &
$u_0=-12 g_x^2$ &
$u_0=-30 g_x^2$ \\
$r\!\!=\!\!-3,\!-1,6,8,10\!$ & 
$\!r\!\!=\!\!-2,\!-1,5,6,\!12\!\!$ &
$r\!\!=\!\!-7,\!-1,6,10,12\!$ &
$\!r\!\!=\!\!-5,\!-1,6,8,\!12\!\!$ \\
OK for $r \ge 6$ &
OK for $r \ge 5$ &
OK for $r \ge 6$ &
OK at $r\!=\!6,8$ \\ 
& & & Not at $r\!=\!12$ \\
\hline 
Passes Test & Passes Test & Passes Test & Fails Test \\
\hline
\end{tabular}
\vskip 2pt
\noindent
Table 1: $\;\;$ Painlev\'e analysis of fifth-order KdV equation 
\newline
$\;\;\;\;\;\;\; u_t + a u^2 u_x + b u_x u_{2x} + c u u_{3x} + u_{5x} = 0 $
\end{center}
\subsection{Simple Systems}
We start with a famous system of ODEs,
\begin{equation}
\label{lorenzsys}
u_1' = a (u_2 - u_1), 
\;\quad 
u_2' = - u_1 u_3 + b u_1 - u_2, 
\;\quad 
u_3' = u_1 u_2 - c u_3,
\end{equation}
where $a,b,$ and $c$ are positive constants. 
System (\ref{lorenzsys}) was proposed by the meteorologist E. N. Lorenz 
as a simplified model for atmospheric turbulence in a vertical air 
cell beneath a thunderhead \cite{sparrow}. 

Using our code {\it painsys.max} or {\it painsys.m}, with
the series (\ref{odeseries}) for $i=1,2,3,$ we determine the 
leading orders $\alpha_1 = -1, \alpha_2 = \alpha_3 = -2.$
The first coefficients in the series are
\begin{equation}
\label{lorenzleading}
u_0^{(1)} =  \pm 2 i, \quad u_0^{(2)} = \mp \frac{2 i}{a}, \quad
u_0^{(3)} = -\frac{2}{a}.
\end{equation}
The roots are $r=-1, 2, 4.$ Furthermore, 
\begin{equation}
\label{lorenzco1}
u_1^{(1)} = \mp \frac{1}{3} i (2 c - 3 a +1), \quad
u_1^{(2)} = \pm 2 i, \quad
u_1^{(3)} = \frac{2}{3 a} (3 a - c + 1).
\end{equation}
The compatibility conditions at resonances $r=2$ and $r=4$ are not
satisfied. At resonance $r=2$ we encounter the condition
$a (c - 2a) (c + 3a - 1) = 0.$ So, we consider two cases:
\begin{itemize}
\item[]
$\bullet$ For $c=2a,$ the compatibility condition at $r=4$ is not satisfied.
\item[] 
$\bullet$ For $c=1-3a,$ the compatibility condition at $r=4$ is satisfied
\item[] $\;\;\;$provided that $a = \frac{1}{3}.$
\end{itemize}
We conclude that the Lorenz system (\ref{lorenzsys}) passes the 
Painlev\'e test when $a=\frac{1}{3}$ and $c=0.$ 
This special case may not be relevant in the context of meteorology since 
in (\ref{lorenzsys}) the parameters $a,b,c$ 
in (\ref{lorenzsys}) are supposed to be positive. 

To illustrate the generalization of the algorithm to systems of PDEs, 
consider the system
% \begin{eqnarray} \label{particle}
\begin{equation} 
\label{particle}
u u_{xt} - u_x u_t + v_x v_t =  0, \quad
u v_{xt} - u_t v_{x}  - u_x v_t  =  0, 
\end{equation}
% \end{eqnarray}
which plays a role in elementary particle physics \cite{tamsah}.
For any of the dependent variables we introduce a Laurent expansion
\begin{equation} 
\label{twolaurent}
u = g^{\alpha} {\sum_{k=0}^{\infty} u_{k} \; g^{k}}, \quad
v = g^{\beta} {\sum_{k=0}^{\infty} v_{k} \; g^{k}}.
\end{equation}
Analysis of the dominant behavior leads to $\alpha = \beta = -1$
and $u_0^2 + v_0^2 = 0 $. 
Let $v_0$ be arbitrary, then $u_0 = \pm i v_0. $
Next, we look for powers of $g$ at which arbitrary functions 
$u_r , v_r $
can enter. Equating the coefficients of $g^{r-4}, r>0,$ we obtain the system
\begin{equation} \label{matrixr}
\left[
\begin{array}{cc} 
 2 r & i r ( r - 1 ) \\
 i ( r^2 - r + 2 ) &  - 2 (r - 1 ) 
\end{array}
\right]
\left[
\begin{array}{c}
u_r \\
v_r 
\end{array}
\right] = 0 .
\end{equation}
The determinant of the coefficient matrix vanishes provided $r=-1, 0,1,2.$ 
Note that $r=0$ confirms that $v_0$ can be chosen freely.
Upon substitution of the Laurent series 
(truncated at level $k=2$) and
setting the terms in $g^{-2}$ and $g^{-3}$ equal to zero, one finds
that $u_1$ is completely determined, whereas $v_1$ and $u_2$ (or 
$v_2 $) are arbitrary. The compatibility conditions are satisfied.
The Laurent expansions have the required number of arbitrary functions
$u_k$. Thus, the system (\ref{particle}) passes the Painlev\'e test.

Hirota and Satsuma \cite{ablcla} proposed a coupled system of KdV equations, 
\begin{equation}
\label{HirSatpar}
u_t - 6 a u u_x + 6 v v_x - a u_{3x} = 0, \;\quad 
v_t + 3 u v_x + v_{3x} = 0,
\end{equation}
where $a$ is a nonzero parameter. 
System (\ref{HirSatpar}) describes interactions of two long waves with 
different dispersion relations. It is known to be completely integrable for 
$a=\frac{1}{2}.$
This is confirmed by the Painlev\'e test. 
Indeed, with (\ref{twolaurent}) we obtain $\alpha = \beta = -2$ and 
$r=-2,-1,3,4,6$ and $8.$
Furthermore, $u_0=-4$ and $v_0=\pm 2 \sqrt{2 a},$ determine the 
coefficients $u_1, v_1, u_2, v_2$ unambiguously. 
At resonances $3$ and $4$ there is one free function and no condition 
for $a.$ The coefficients $u_5$ and $v_5$ are unique determined, 
but at resonance $6,$ the compatibility condition is only satisfied 
if $a = \frac{1}{2}.$ For this value, the compatibility condition 
at resonance $8$ is also satisfied. 

For the integrable version of the Boussinesq system \cite{sac}
% \begin{eqnarray} 
\begin{equation}
\label{boussystem}
u_t + v_x + u u_x = 0, \quad\quad
v_t + u_{3x} + (uv)_x = 0, 
\end{equation}
% \end{eqnarray}
with the Laurent expansions in (\ref{twolaurent}), 
the leading order is $\alpha = -1, \beta = -2.$
Careful investigation of the recursion relations linking the $u_k$ and
$v_k$ allows one to conclude that there are resonances at levels 
$2, 3$ and $4$. Finally, the three compatibility conditions are seen 
to hold after lengthy computations. System (\ref{boussystem}) thus 
passes the test. 

Our codes {\it painsys.max} are also applicable to complex equations 
such as the nonlinear Schr\"odinger (NLS) equation \cite{ablcla}, 
\begin{equation}
\label{nlspainleve}
i q_t - q_{2x} + 2 {|q|}^2 q =0,
\end{equation}
where $q(x,t)$ is a complex function. 
First, (\ref{nlspainleve}) must be rewritten as a system:
\begin{equation}
\label{NLSsys}
u_t - v_{2x} + 2 v (u^2+v^2) = 0, \quad\quad
v_t + u_{2x} - 2 u (u^2+v^2) = 0,
\end{equation}
where $q = u + i v.$ 
For simplicity, we use the Kruskal simplification, setting 
$g(x,t) = x - h(t)$ in (\ref{twolaurent}).
Then, the leading order is $\alpha = \beta = -1,$ 
and $r = -1, 0, 3 $ and $4.$
Furthermore, $v_0 = \pm \sqrt{1 - u_0^2},$ with $u_0$ arbitrary, and
\begin{eqnarray}
\label{coefnls}
u_1 &=& \frac{1}{2} v_0 h_t, \quad\;
v_1 = -\frac{1}{2} u_0 h_t, \quad\;
u_2 = - \frac{1}{12 v_0} u_0 (v_0 h_t^2 + 2 u_{0,t}), 
\nonumber \\
v_2 &=& -\frac{1}{12} (v_0 h_t^2 + 2 u_{0,t}), \quad\;
u_3 = - \frac{1}{8 u_0} (h_{tt} + 8 v_0 v_3), 
\nonumber \\
u_4 &=& - \frac{1}{24 v_0^4} (2 v_0^3 h_t h_{tt} - v_0^2 u_{0,tt} 
     - u_0 u_{0,t}^2 - 24 u_0 v_0^3 v_4),
\end{eqnarray}
where $v_3$ and $v_4$ are arbitrary. 
Note that at every resonance there is one free function in the 
Laurent expansion. The NLS equation passes the test.

% Finally, we test the three-component extension of the KdV equation, 
% \begin{eqnarray}
% \label{3cKdV}
% u_t -6 u u_x +2 v v_x+2 w w_x-u_{3x} &=& 0, \nonumber \\
% v_t-2 v u_x-2 u v_x &=& 0, \\
% w_t -2 w u_x-2 u w_x &=& 0, \nonumber
% \end{eqnarray}
% due to Kupershmidt \cite{kuper}.
% System (\ref{3cKdV}) can be written as a bi-Hamiltonian system with 
% infinitely many conservation laws \cite{jsc}. 
% Using 
% \begin{equation} 
% \label{threelaurent}
% u  =   g^{\alpha} {\sum_{k=0}^{\infty} u_{k} \; g^{k}}, \quad
% v  =  g^{\beta} {\sum_{k=0}^{\infty} v_{k} \; g^{k}}, \quad
% w  =  g^{\gamma} {\sum_{k=0}^{\infty} w_{k} \; g^{k}}.
% \end{equation}
% we determine the leading order behavior ......

In searching the literature we found numerous examples of systems for 
which the Painlev\'e test was carried out by hand.
Some of the most elaborate examples of Painlev\'e analysis 
involve three-wave interactions \cite{menchelee}, the mixmaster universe 
model \cite{congraram}, and chaotic star pulsations \cite{verherser}. 

\section{Symbolic Programs for Conserved Densities and Symmetries}
\subsection{The Key Concepts}
The key observation behind our algorithms is that conserved densities and 
higher-order symmetries of a PDE or DDE system abide by the dilation 
symmetry of that system. We illustrate this for prototypical examples 
of single PDEs and DDEs. 

{\bf Dilation Invariance of PDEs.}$\;\;$The ubiquitous Korteweg-de Vries 
(KdV) equation from soliton theory \cite{ablcla}, 
\begin{equation}
\label{kdv}
u_t = 6 u u_x + u_{3x},
\end{equation} 
is invariant under the dilation (scaling) symmetry
\begin{equation}
\label{kdvdilation}
(t, x, u) \rightarrow ({\lambda}^{-3} t, \lambda^{-1} x, {\lambda}^{2} u),
\end{equation}
where $\lambda$ is an arbitrary parameter. 
Obviously, $u$ corresponds to two derivatives in $x$, 
i.e.\ $u \sim {\partial^2}/{\partial {x^2}}. $
Similarly, $ {\partial}/{\partial t} \sim {\partial}^3/{\partial x}^3.$ 
Introducing weights, denoted by $w,$ we have $w(u) = 2$ and 
$w({\partial}/{\partial t})=3,$ provided we set 
$w({\partial}/{\partial x}) = 1.$ 
The {\it rank} $R$ of a monomial equals the sum of all of its weights. 
Observe that (\ref{kdv}) is {\it uniform in rank} since all the terms
have rank $R=5$. 
\vskip 4pt
\noindent
{\bf Conserved Densities of PDEs.}$\;\;$For PDEs like (\ref{kdv}), 
the conservation law 
\begin{equation}
\label{PDEconslaw}
{\rm D}_{t} \rho + {\rm D}_{x} J = 0 
\end{equation}
connects the {\rm conserved density\/} $\rho$ and the associated 
{\em flux\/} $J.$ 
As usual, ${\rm D}_{t}$ and ${\rm D}_{x}$ are total derivatives.
With few exceptions, {\it polynomial} density-flux pairs only 
depend on ${\bf u}, {\bf u}_x, $ etc., and not explicitly on $t$ and $x.$

The first three (of infinitely many) independent conservation laws 
for (\ref{kdv}) are 
\begin{eqnarray} 
\label{kdvconslaw1and2}
& & {\rm D}_t (u)  - {\rm D}_x (3 u^2 + u_{2x} )=0,\quad 
{\rm D}_t (u^2) - {\rm D}_x ( 4 u^3 - u_x^2 + 2 u u_{2x} )=0, \\
\label{kdvconslaw3}
& & {\rm D}_t (u^3 - \frac{1}{2} u_x^2) -
{\rm D}_x ({9 \over 2} u^4 - 6 u u_x^2 + 3 u^2 u_{2x} 
+ \frac{1}{2} u_{2x}^2 - u_x u_{3x})=0.
\end{eqnarray}
The densities $\rho = u, u^2, u^3 - \frac{1}{2} u_x^2 $ have 
ranks $2, 4$ and $6$, respectively. The associated fluxes have ranks
$4, 6$ and $8.$ The terms in the conservation laws have ranks 
$5, 7,$ and $9.$

Integration of both terms in the conservation law with respect to $x$ yields 
\begin{equation}
\label{conservedP}
P = \int_{-\infty}^{+\infty} \rho \; dx = {\rm constant \; in \; time},
\end{equation}
provided $J$ vanishes at infinity. $P$ is the true conserved quantity. 
The first two conservation laws correspond to conservation of 
momentum and energy. For ODEs, the quantities $P$ are called constants 
of motion. 
\vskip 4pt
\noindent
{\bf Symmetries of PDEs.}$\;\;$
As summarized in Table 2, 
${\bf G} (x, t, {\bf u}, {\bf u}_{x}, {\bf u}_{2x}, ...)$
is a {\it symmetry\/} of a PDE system if and only if it leaves it invariant for 
the change ${\bf u} \rightarrow {\bf u} + \epsilon {\bf G}$ within order 
$\epsilon.$ Hence, 
$ {\rm D}_t ({\bf u} + \epsilon {\bf G}) = 
{\bf F} ({\bf u} + \epsilon {\bf G}) $
must hold up to order $\epsilon.$ 
Thus, ${\bf G}$ must satisfy the linearized equation 
$ {\rm D}_t {\bf G} = {\bf F}'({\bf u})[{\bf G}], $ 
where ${\bf F}'$ is the Fr\'echet derivative:
${\bf F}'({\bf u})[{\bf G}] = {\partial \over \partial{\epsilon}} 
{\bf F}({\bf u}+\epsilon {\bf G}) |_{\epsilon = 0}.$
\vskip 0.01pt
\begin{center}
% Start Table 2 (Definitions) 
% \begin{table}[hp] \caption{{\rm Conservation Laws and Symmetries}}
% \vskip .001pt
% \noindent
% \begin{footnotesize}
\begin{tabular}{ | l | l | l |} \hline 
& Continuous $\!$Case $\!$(PDEs) & Semi-discrete Case (DDEs) \\ [0.5ex] 
\hline
% & & \\
System & 
${\bf u}_t={\bf F}({\bf u}, {\bf u}_{x}, {\bf u}_{2x}, ...)$ &
${\dot{\bf u}}_n \!=\!{\bf F}(...,
{\bf u}_{n-1}, {\bf u}_{n}, {\bf u}_{n+1},...)$ \\
% & & \\
\hline 
% & & \\
Conservation Law & $ {\rm D}_{t} \rho + {\rm D}_{x} J = 0 $ &
${\dot{\rho}}_n + J_{n+1} - J_n = 0 $ \\
% & & \\
\hline 
% & & \\
Symmetry & ${\rm D}_t{\bf G} = {\bf F}'({\bf u})[{\bf G}] $
& $ {\rm D}_t{\bf G}={\bf F}'({\bf u}_n)[{\bf G}] $ \\
& $\;\;\;\;\;\;\;$ 
$={\partial\over\partial{\epsilon}}{\bf F}({\bf u}
+\epsilon {\bf G})|_{\epsilon=0} $ 
& $\;\;\;\;\;\;\;$ 
$ = {\partial \over \partial{\epsilon}}{\bf F}({\bf u}_n 
+\epsilon {\bf G})|_{\epsilon=0}$ \\
% & & \\
\hline 
\end{tabular}
\vskip 2pt
\noindent
Table 2: $\;\;$ Conservation Laws and Symmetries
\end{center}
% \end{footnotesize}
% \end{table}
\vfill
\newpage
% \vskip 0.01pt
\noindent
Two nontrivial higher-order symmetries \cite{PO1993} 
of (\ref{kdv}) are
\begin{eqnarray}
\label{twosymm1}
G^{(1)} \!&=&\! 30 u^2 u_x + 20 u_x u_{2x} + 10 u u_{3x} + u_{5x}, \\
\label{twosymm2}
G^{(2)} \!&=&\! 140 u^3 u_x + 70 u_x^3 + 280 u u_x u_{2x} + 70 u^2 u_{3x} 
+ 70 u_{2x} u_{3x} + 42 u_x u_{4x} \nonumber \\
&&  + 14 u u_{5x} + u_{7x}.
\end{eqnarray}
The recursion operator \cite[p. 312]{PO1993}, 
\begin{equation}
\label{recursion}
\Phi = D^2 + 4 u + 2 u_x D^{-1},
\end{equation}
connects the above symmetries, $ \Phi \, G^{(1)} = G^{(2)},$ and also the 
lower order symmetries as shown in \cite{baltzer}. 

Note that the recursion operator (\ref{recursion}) is also uniform in 
rank with $R=2$ since $w(D^{-1}) = -1.$ Currently, we are working on 
an algorithm and symbolic code to compute recursion operators 
based on the knowledge of a few higher-order symmetries and conserved
densities.

Higher-order symmetries lead to new integrable PDEs. 
Indeed, the evolution equations $u_t = G^{(1)} $ 
and $u_t = G^{(2)} $ are the well-known fifth and seventh-order equations
in the completely integrable KdV hierarchy \cite{ablcla}. 

{\bf Dilation Invariance of DDEs.}$\;\;$We now turn to the semi-discrete 
case. As prototype, consider the Volterra lattice \cite{ablcla},  
\begin{equation} 
\label{volterra}
{\dot{u}}_n = u_n \, (u_{n+1}-u_{n-1}). 
\end{equation}
which is one of the discretizations of (\ref{kdv}).
Note that (\ref{volterra}) is invariant under
\begin{equation}
\label{volterradilation}
(t, u_n) \rightarrow (\lambda^{-1} t, \lambda u_n).
\end{equation}
So, $u_n \sim {\rm d}/{\rm dt},$ or $w(u_n)=1$ if we set 
$w({\rm d}/{\rm dt})=1.$ 
Every term in (\ref{volterra}) has rank $R=2,$ thus (\ref{volterra}) 
is uniform in rank.
\vskip 4pt
\noindent
{\bf Conserved Densities of DDEs.}$\;\;$For DDEs like (\ref{volterra}), the 
conservation law 
\begin{equation}
\label{DDEconslaw}
{\dot{\rho}}_n + J_{n+1} - J_n = 0 
\end{equation}
connects the {\rm conserved density\/} $\rho_n$ and the associated 
{\em flux\/} $J_n.$ 
For (\ref{volterra}) the first two conservation laws 
(of ranks $2$ and $3$) are
\begin{eqnarray}
\label{volterraconslaw3}
\!\!\!\!\!\!\!\! & & \!\!\frac{d}{dt}(u_n) 
+ u_n u_{n-1} - u_{n+1} u_n =0,  \\
\!\!\!\!\!\!\!\! & & \!\! \frac{d}{dt}(\frac{1}{2} u_n^2 + u_n u_{n+1}) 
+ u_{n-1} u_n^2 + u_{n-1} u_n u_{n+1} - u_n u_{n+1}^2 - u_n u_{n+1} u_{n+2}=0. 
\end{eqnarray}
The densities $\rho_n = u_n$ and 
$\rho_n = u_n \, (\frac{1}{2} u_n + u_{n+1}) $
have ranks $1$ and $2.$ Their fluxes $J_n = - u_n u_{n-1}$ 
and $J_n = - u_{n-1} u_n^2 - u_{n-1} u_n u_{n+1} $ have ranks $2$ and $3.$
\vskip 4pt
\noindent
{\bf Symmetries of DDEs.}$\;\;$For DDEs of type (\ref{volterra}), 
${\bf G} (...,{\bf u}_{n-1}, {\bf u}_{n}, {\bf u}_{n+1},...)$ 
is a {\it symmetry\/} if and only if the infinitesimal transformation
$\; {\bf u}_n \rightarrow {\bf u}_n + \epsilon {\bf G}$ leaves 
the DDE invariant within order $\epsilon.$ Consequently, ${\bf G}$ 
must satisfy ${{\rm d}{\bf G} \over {\rm d}{t}} = 
{\bf F}'({\bf u}_n)[{\bf G}], $ where ${\bf F}'$ is the Fr\'echet derivative,
$ {\bf F}'({\bf u}_n)[{\bf G}] = {\partial \over \partial{\epsilon}} 
{\bf F}({\bf u}_n +\epsilon {\bf G}) |_{\epsilon = 0}.$ See Table 2.

The first nontrivial higher-order symmetry of (\ref{volterra}) is 
\begin{equation}
\label{volterrasymm}
G = u_n u_{n+1} \, (u_n + u_{n+1} + u_{n+2}) - u_{n-1} u_n \,
(u_{n-2} + u_{n-1} + u_n),
\end{equation}
and, similar to the continuous case, an integrable lattice 
${\dot u_n} = G$ follows. 

Both (\ref{kdv}) and (\ref{volterra}) have infinitely many polynomial 
conserved densities \cite{jsc,physletta} and symmetries \cite{baltzer}.
\vskip 1pt
% \vfill
% \newpage
\noindent
{\it Remarks:}
\vskip 1pt
\noindent
(i) For scaling invariant systems like (\ref{kdv}) 
and (\ref{volterra}), it suffices to consider the dilation symmetry 
on the space of independent and dependent variables. 

(ii) For systems that are inhomogeneous under a suitable scaling 
symmetry we use the following trick: 
We introduce one (or more) auxiliary parameter(s) with appropriate scaling. 
In other words, we extend the action of the dilation symmetry to the space of 
independent and dependent variables, {\it including} the parameters.
These extra parameters should be viewed as additional dependent variables, 
with the caveat that their derivatives are zero. 
With this trick we can apply our algorithms to a larger class of 
polynomial PDE and DDE systems. Examples can be found in 
\cite{jsc,physicad,baltzer,physletta}.

Scaling (dilation) invariance, which is a special Lie-point symmetry, is 
common to many integrable nonlinear PDEs and DDEs.
In the next two sections we show how the scaling invariance can be
explicitly used to compute polynomial conserved densities and 
higher-order symmetries of PDEs and DDEs. 
\subsection{Algorithms for the Computation of Conserved Densities}
{\bf PDE Case.}$\;$Conserved densities of PDEs can be computed as follows:
\vskip 1pt
\noindent
$\bullet$ Require that each equation in the system of PDEs is uniform 
in rank. Solve the resulting linear system to determine the weights of 
the dependent variables. 
For instance for (\ref{kdv}), solve 
$ w(u) + w({\partial}/{\partial t}) = 2 w(u) + 1 = w(u) + 3, $ to get 
$ w(u) = 2$ and $w({\partial}/{\partial t}) = 3.$ 
\vskip 1pt
\noindent
$\bullet$
Select the rank $R$ of $\rho ,$ say, $R=6.$ 
Make a linear combination of all the monomials in the components of 
${\bf u}$ and their $x$-derivatives that have rank $R.$ 
Remove all monomials that  are total $x$-derivatives (like $u_{4x}$).
Remove equivalent monomials, that is, those that only differ by a 
total $x$-derivative. For example, $u u_{2x} $ and $u_x^2$ are equivalent 
since $u u_{2x} = \frac{1}{2} (u^2)_{2x} - u_x^2.$ 
For (\ref{kdv}), one gets 
$\rho = c_1 u^3 + c_2 u_x^2$ of rank $R=6.$
\vskip 1pt
\noindent
$\bullet$
Substitute $\rho$ into the conservation law 
(\ref{PDEconslaw}). Use the PDE system to eliminate all $t$-derivatives, 
and require that the resulting expression $E$ is a total $x$-derivative. 
Apply the Euler operator (see \cite{jsc} for the general form),
\begin{equation}
\label{euleroperator}
{\cal L\/}_{u} = \frac{\partial }{\partial{u}} -
D_x (\frac{\partial}{\partial{{u_x}}})+
D_{x}^2 ( \frac{\partial }{\partial{{u_{2x}}}})+\cdots+
(-1)^n D_{x}^n (\frac{\partial }{\partial{u_{nx}}})
\end{equation}
to $E$ to avoid integration by parts. 
If any terms remain, they must vanish identically. 
This yields a linear system for the constants $c_i.$ Solve the system. 
For (\ref{kdv}), one gets $c_1 = 1, c_2 = -1/2.$
\vskip 1pt
\noindent
See \cite{jsc} for the complete algorithm and its implementation.
\vskip 3pt
\noindent
{\bf DDE Case.}$\;$The computation of conserved densities proceeds 
as follows:  
\vskip 1pt
\noindent
$\bullet$ Compute the weights in the same way as for PDEs. 
For (\ref{volterra}), one gets $w(u_n) =1 $ by solving 
$w(u_n) + 1 = 2 w(u_n).$ 
Note that $w({\rm d}/{\rm dt})=1$ and weights are independent of $n.$
\vskip 1pt
\noindent
$\bullet$ Determine all monomials of rank $R$ in the components of 
${\bf u}_n$ and their $t$-derivatives. 
Use the DDE to replace all the $t$-derivatives.
Monomials are {\it equivalent} if they belong to the same equivalence class
of shifted monomials. For example, $u_{n-1} v_{n+1}, 
u_{n+2} v_{n+4}$ and $u_{n-3} v_{n-1}$ are equivalent. 
Keep only the main representatives (centered at $n$) of the various classes.
\vskip 1pt
\noindent
$\bullet$ Combine these representatives linearly with coefficients $c_i,$
and substitute the form of $\rho_n$ into the conservation law 
$\dot \rho_n = J_n - J_{n+1}. $
\vskip 1pt
\noindent
$\bullet$ Remove all $t$-derivatives and pattern-match the resulting 
expression with $J_n - J_{n+1}.$ 
To do so use the following {\it equivalence criterion}: if two monomials
$m_1$ and $m_2$ are equivalent, $m_1 \, \equiv m_2,$ then
$m_1 \,=\, m_2 + [M_n - M_{n+1} ] $ for some polynomial $M_n$ 
that depends on ${\bf u}_n$ and its shifts.
For example, $ u_{n-2} u_n \, \equiv \, u_{n-1} u_{n+1} $
since $ u_{n-2} u_n = u_{n-1} u_{n+1} + [u_{n-2} u_{n} - u_{n-1} u_{n+1}] 
= u_{n-1} u_{n+1} + [M_n - M_{n+1}], $ with $M_n = u_{n-2} u_{n}.$
Set the non-matching part equal to zero, and solve the linear system 
for the $c_i.$ Determine $J_n$ from the pattern $J_n - J_{n+1}.$ 
For (\ref{volterra}), the first three (of infinitely many) densities 
$\rho_n$ are listed in Table 3. 
\vskip 0.02pt
\noindent
\begin{center}
% Start table 3 (Leading Examples) 
% \begin{table}[hp] \caption{{\rm Prototypical Examples}}
% \vskip 2pt
% \noindent
\begin{tabular}{ | l | l | l |} \hline 
& KdV Equation & Volterra Lattice \\ [0.5ex] 
\hline
% & & \\
Equation & 
$u_t = 6 u u_x + u_{3x}$ &
${\dot{u}}_n = u_n \, (u_{n+1}-u_{n-1})$\\
% & & \\
\hline 
% & & \\
Densities & $\rho = u, \;\; \rho = u^2 $ & ${\rho}_n = u_n, \;\;
{\rho}_n = u_n \, (\frac{1}{2} u_n + u_{n+1})$ \\
& $\rho = u^3 - \frac{1}{2} u_x^2$ & 
${\rho}_n \!=\! \frac{1}{3} u_n^3 \!+ 
u_n u_{n+1} (u_n \!+ u_{n+1} \!+ u_{n+2})\!$ \\
% & & \\
\hline 
% & & \\
$\!$Symmetries$\!$ & $G\!=\!u_x,\;\;G\!=\!6 u u_x + u_{3x} $ 
&$G = u_n u_{n+1} \, (u_n + u_{n+1} + u_{n+2}) $ \\ 
&$ G \!=\!30 u^2 u_x + 20 u_x u_{2x}$  &
$ \;\;\;\;\;\;\;\; - u_{n-1} u_n  (u_{n-2} + u_{n-1} + u_n)$ \\
& $\;\;\;\;\;\;\; + 10 u u_{3x} + u_{5x}$ & \\
% & & \\
\hline 
\end{tabular}
% \end{table}
\vskip 2pt
\noindent
Table 3: $\;\;$ Prototypical Examples
\end{center}
% \end{footnotesize}
% \end{table}
\vskip 0.01pt
\noindent
Details about this algorithm and its implementation are in 
\cite{physicad,physletta}. 
See \cite{invsymmsoft} for an integrated {\em Mathematica} Package that 
computes conserved densities (and also symmetries) of PDEs and DDEs. 

\subsection{Algorithm for the Computation of Symmetries}

{\bf PDE Case.}$\;$ Higher-order (or generalized symmetries) of PDEs 
can be computed as follows:
\vskip 1pt
\noindent
$\bullet$ Determine the weights of the dependent variables in the system. 
\vskip 1pt
\noindent
$\bullet$ Select the rank $R$ of the symmetry. 
Make a linear combination of all the monomials involving ${\bf u}$ and
its $x$-derivatives of rank $R.$ 
For example, for (\ref{kdv}), 
$G = c_1 \, u^2 u_x + c_2 \, u_x u_{2x} + c_3 \, u u_{3x} + c_4 \, u_{5x}$
is the form of the generalized symmetry of rank $R=7.$ 
In contrast to the computation of conserved densities, 
no terms are removed here.
\vskip 1pt
\noindent
$\bullet$ Compute ${\rm D}_t {\bf G}.$ 
Use the PDE system to remove all $t$-derivatives. 
Equate the result to the Fr\'echet derivative ${\bf F}'({\bf u})[{\bf G}].$
Treat the different monomial terms in ${\bf u}$ and its $x$-derivatives 
as independent, to get the linear system for $c_i.$ Solve that system. 
For (\ref{kdv}), one obtains
\begin{equation}
\label{laxsymmagain}
G =  30 u^2 u_x + 20 u_x u_{2x} + 10 u u_{3x} + u_{5x}. 
\end{equation}
The symmetries of the Lax family of rank 3, 5, and 7 are listed in 
Table 3. They are the first three of infinitely many. 
See \cite{baltzer} for the details about the algorithm.
% \vskip 2pt
\vfill
\newpage
\noindent
{\bf DDE Case.}$\;$Higher-order symmetries of DDEs can be computed 
as follows: 
\vskip 1pt
\noindent
$\bullet$ First determine the weights of the variables in the DDE
the same way as for conserved densities.
\vskip 1pt
\noindent
$\bullet$ Determine all monomials of rank $R$ in the components of 
${\bf u}_n$ and their $t$-derivatives. 
Use the DDE to replace all the $t$-derivatives.
Make a linear combination of the resulting monomials with coefficients $c_i.$
\vskip 1pt
\noindent
$\bullet$ 
Compute $D_t {\bf G}$ and remove all $ {\dot {\bf u}}_{n-1}, 
{\dot {\bf u}}_n, {\dot {\bf u}}_{n+1},$ etc. 
Equate the resulting expression to the Fr\'echet derivative 
$ {\bf F}'({\bf u}_n)[{\bf G}] $ and solve the system for the $c_i,$
treating the monomials in ${\bf u}_n$ and its shifts as independent. 

For (\ref{volterra}), the symmetry ${\bf G}$ of rank $R=3$ is listed 
in Table 3. There are infinitely many symmetries, all with different ranks.

See \cite{baltzer} for the complete algorithm and its implementation in 
{\it Mathematica}, 
and \cite{invsymmsoft} for an integrated {\em Mathematica} Package that
computes symmetries of PDEs and DDEs.
\vskip .4pt
\noindent
{\it Notes:}
\vskip .4pt
\noindent
(i) A slight modification of these methods allows one to find 
conserved densities and symmetries of PDEs that explicitly depend 
on $t$ and $x.$ See the first example in the next section.
\vskip 1pt
\noindent
(ii) Applied to systems with free parameters, the linear system for the 
$c_i$ will depend on these parameters. A careful analysis of the eliminant 
leads to conditions on these parameters so that a sequence of conserved
densities or symmetries exists. 
Details about this type of analysis and its computer implementation 
can be found in \cite{jsc}. 
\section{Examples of Densities and Symmetries}

\subsection{Vector Modified KdV Equation}
In \cite{verheest}, Verheest investigated the integrability of
a vector form of the modified KdV equation (vmKdV), 
\begin{equation} 
\label{vectormkdv}
{\bf B}_t + (|{\bf B}|^2 {\bf B})_x + {\bf B}_{xxx} = 0,
\end{equation}
or component-wise for ${\bf B}=(u,v),$
\begin{eqnarray}
\label{vmkdv}
u_t + 3 u^2 u_x + v^2 u_x + 2 u v v_x + u_{3x}  = 0, 
\nonumber \\
v_t + 3 v^2 v_x + u^2 v_x + 2 u v u_x + v_{3x} = 0.
\end{eqnarray}
With our software {\it InvariantsSymmetries.m} \cite{invsymmsoft} 
we computed
\begin{eqnarray}
\rho_1 &=& u, \quad \rho_2 = v, \quad \rho_3 = u^2 + v^2, \quad 
\rho_4 = \frac{1}{2} (u^2 + v^2)^2 - (u_x^2 +  v_x^2),  \\
\rho_5 &=& 
\frac{1}{3} x (u^2 + v^2) - \frac{1}{2} t (u^2 + v^2)^2 + t (u_x^2 + v_x^2).
\end{eqnarray}
Note that the latter density depends explicitly on $x$ and $t.$
Verheest \cite{verheest} has shown that (\ref{vmkdv}) is non-integrable
for it lacks a bi-Hamiltonian structure and recursion operator. 
We were unable to find additional polynomial conserved densities. 
Polynomial higher-order symmetries for (\ref{vmkdv}) do not appear to exist.

\subsection{Heisenberg Spin Model} 
The continuous Heisenberg spin system \cite{fadtak} or Landau-Lifshitz 
equation,
\begin{equation}
\label{orgheisenberg}
{\bf S}_t = {\bf S} \times \Delta {\bf S} + {\bf S} \times D {\bf S},
\end{equation}
models a continuous anisotropic Heisenberg ferromagnet. 
It is considered a universal integrable system since various known
integrable PDEs, such as the NLS and sine-Gordon equations, can be
derived from it. 
In (\ref{orgheisenberg}), ${\bf S} = [u,v,w]^T $ with real components,
$\Delta = \nabla^2$ is the Laplacian, $D$ is a diagonal matrix,
and $\times$ is the standard cross product of vectors.

Split into components, (\ref{orgheisenberg}) reads
% \begin{equation}
\begin{eqnarray}
\label{orgheisenbergcomponents}
u_t &=& v w_{2x} - w v_{2x} + (\beta - \alpha) v w, \nonumber \\
v_t &=& w u_{2x} - u w_{2x} + (1 - \beta) u w, \nonumber \\
w_t &=& u v_{2x} - v u_{2x} + (\alpha - 1) u v.
\end{eqnarray}
% \end{equation}
% \vfill
% \newpage
% \noindent
In Table 4 we list the conserved densities for three typical cases; 
other cases are similar. 

Note that for all the cases we considered
\begin{equation}
\label{Ssquare}
\rho = u^2 + v^2 + w^2 = ||{\bf S}||^2 
\end{equation}
is constant in time (since $J = 0).$ 
Hence, all even powers of $||{\bf S}||$ are also conserved densities, 
but they are dependent of (\ref{Ssquare}). 
\vskip 0.01pt
\noindent
\begin{center}
% Start Table 4 (Heisenberg continuous case only) 
% \begin{table}[hp] \caption{{\rm Conserved Densities for the continuous
% Heisenberg Spin Model}}
% \vskip .001pt
% \noindent
% \begin{footnotesize}
\begin{tabular}{ | l | l | l|} \hline 
$D=diag(1,\alpha,\beta)$ & $D=diag(1,\alpha,0)$ & $D=diag(0,0,0)$ \\
$\alpha\ne 0, \beta \ne 0$ & $ \alpha\ne 0$ & \\ [0.5ex] 
\hline
% & & \\
$\rho =u$ if $\alpha=\beta $ & % C1
$\rho=w$ if $\alpha=1 $      & % C2
$\rho= u$ \\ % C3 
$\rho =v$ if $\beta=1$  & % C1
$\rho = u^2 + v^2 + w^2 $   & % C2
$\rho= v$ \\ % C3 
$\rho =w$ if $\alpha=1$ & % C1
$\rho =(1-\alpha) v^2 + w^2$ & % C2
$\rho= w$ \\ % C3 
$\rho = u^2 + v^2 + w^2 $ & % C1
$\;\;\;\;\;\; + u_x^2 + v_x^2 + w_x^2 $ & % C2 
$\rho = u^2 + v^2 + w^2 $ \\ % C3 
$\rho = (1-\alpha) v^2 + (1 -\beta) w^2$ & % C1 
& % C2 
$\rho = u_x^2 + v_x^2 + w_x^2 $  \\ % C3 
$\;\;\;\;\;\;\; + u_x^2 + v_x^2 + w_x^2 $ & % C1
& % C2 
\\ % C3
\hline 
\end{tabular}
\vskip 2pt
\noindent
Table 4: $\;\;$ Conserved Densities for the Heisenberg Spin Model
\end{center}
% \end{footnotesize}
% \end{table}
% \vfill
% \newpage
\vskip 0.01pt
\noindent
Furthermore, the sum of two conserved densities is a conserved density. 
Hence, after adding (\ref{Ssquare}),
\begin{equation}
\label{contspinhamiltonian1}
\rho = (\alpha - 1) v^2 + (\beta -1) w^2 - (u_x^2 + v_x^2 + w_x^2)
\end{equation}
can be replaced by  
\begin{equation}
\label{contspinhamiltonian2}
\rho = u^2 + \alpha v^2 + \beta w^2 - (u_x^2 + v_x^2 + w_x^2).
\end{equation}
Note that $u_x^2 = D_x (u u_x) - u u_{2x}$ and recall that densities are 
equivalent if they only differ by a total $x-$derivative.
So, (\ref{contspinhamiltonian2}) is equivalent with 
\begin{equation}
\label{contspinhamiltonian3}
\rho = u^2 + \alpha v^2 + \beta w^2 + u u_{2x}  + v v_{2x} + w_{2x}, 
\end{equation}
which can be compactly written as 
$\rho = {\bf S} \cdot \Delta {\bf S} + {\bf S} \cdot D {\bf S}, $
where $D=diag(1,\alpha,\beta).$
Consequently, the Hamiltonian of (\ref{orgheisenberg}) 
\begin{equation}
\label{truecondhamiltonian}
{\cal H}
= -\frac{1}{2}\int {\bf S}\cdot\Delta {\bf S}+{\bf S}\cdot D {\bf S} \;\; dx
\end{equation}
is constant in time. The dot $(\cdot)$ refers to the standard inner product
of vectors.

\subsection{Extended Lotka-Volterra and Relativistic Toda Lattices}

Itoh \cite{itoh} studied this extended version of 
the Lotka-Volterra equation (\ref{volterra}):
\begin{equation}
\label{extvolterra}
{\dot{u}}_n = \sum_{r=1}^{k-1} (u_{n-r} - u_{n+r}) u_n.
\end{equation}
For $k=2,$ (\ref{extvolterra}) is (\ref{volterra}), for which three
conserved densities and one symmetry are listed in Table 3. 
In \cite{physicad}, we gave two additional densities and 
in \cite{baltzer} we listed two more symmetries. 
\vskip 2pt
\noindent
For (\ref{extvolterra}), we computed 5 densities and 2 higher-order 
symmetries for $k=3$ through $k=5.$ Here is a partial list of the results: 
\vskip 0.01pt
\noindent
For $k = 3:$
\begin{eqnarray}
\rho_1 &=& u_n, \quad\quad\quad
\rho_2 = \frac{1}{2} u_n^2 + u_n (u_{n+1} + u_{n+2}), \\
\rho_3 &=& \frac{1}{3} u_n^3 + u_n^2 (u_{n+1} + u_{n+2})
+ u_n (u_{n+1} + u_{n+2})^2 \nonumber \\ 
& & + u_n (u_{n+1} u_{n+3} + u_{n+2} u_{n+3} + u_{n+2} u_{n+4}), \\
& & \nonumber \\
G &=& u_n^2 (u_{n+1} + u_{n+2} - u_{n-2} - u_{n-1}) 
+ u_n [(u_{n+1} + u_{n+2})^2 
\nonumber \\
& & - (u_{n-2} + u_{n-1})^2 ] 
+ u_n [u_{n+1} u_{n+3} + u_{n+2} u_{n+3} + u_{n+2} u_{n+4} 
\nonumber \\
& & - (u_{n-4} u_{n-2} + u_{n-3} u_{n-2} + u_{n-3} u_{n-1})]. 
\end{eqnarray}
\noindent
For $k = 4:$
\begin{eqnarray}
\rho_1 &=& u_n, \quad\quad\quad
\rho_2 = \frac{1}{2} u_n^2 + u_n (u_{n+1} + u_{n+2} + u_{n+3}), \\
\rho_3 &=& \frac{1}{3} u_n^3 + u_n^2 (u_{n+1} + u_{n+2} 
+ u_{n+3}) + u_n (u_{n+1} + u_{n+2} + u_{n+3})^2 
\nonumber \\
& & + u_n (u_{n+1} u_{n+4} + u_{n+2} u_{n+4} + u_{n+3} u_{n+4} 
+ u_{n+2} u_{n+5} \nonumber \\
& & + u_{n+3} u_{n+5} + u_{n+3} u_{n+6}), \\
& & \nonumber \\
G & = & u_n [
u_{n+1} u_{n+4} + u_{n+2} u_{n+4} + u_{n+3} u_{n+4} + u_{n+2} u_{n+5} 
\nonumber \\
& & + u_{n+3} u_{n+5} + u_{n+3} u_{n+6} 
- (u_{n-6} u_{n-3} + u_{n-5} u_{n-3} + u_{n-4} u_{n-3}
\nonumber \\
& & + u_{n-5} u_{n-2} - u_{n-4} u_{n-2} + u_{n-4} u_{n-1}) ] + u_n 
[(u_{n+1} + u_{n+2} + u_{n+3})^2 \nonumber \\
& & - u_n (u_{n-3} + u_{n-2} + u_{n-1})^2] 
+ u_n^2 [ u_{n+1} + u_{n+2} + u_{n+3}
\nonumber \\
& &  - (u_{n-3} + u_{n-2} + u_{n-1})].
\end{eqnarray}
% \vfill
% \newpage
% \noindent
Our last example involves the integrable relativistic Toda lattice 
\cite{nunmar}:
\begin{equation}\label{reltodalatt}
{\dot{u}}_n = u_n \; (v_{n+1} - v_n + u_{n+1} - u_{n-1}), \quad
{\dot{v}}_n = v_n \; (u_n - u_{n-1}).
\end{equation} 
We computed the densities of rank $1$ through $5.$ The first three are
\begin{eqnarray}
\rho_1 &=& u_n + v_n, \quad\;\;\;
\rho_2 = {\textstyle \frac{1}{2}} (u_n^2 +v_n^2) + 
u_n (u_{n+1} + v_n + v_{n+1}), \\
\rho_3 &=& {\textstyle \frac{1}{3}} (u_n^3 + v_n^3 ) 
+ u_n^2 (u_{n+1} + v_n + v_{n+1}) + u_n [(u_{n+1} + v_{n+1})^2 
\nonumber \\
&& 
+ u_{n+1} u_{n+2} + u_{n+1} v_n + u_{n+1} v_{n+2} + v_n v_{n+1} + v_n^2].
\end{eqnarray}
We computed the symmetries for ranks $(2,2)$ through $(4,4).$
The first two are: 
\begin{eqnarray}
G_1^{(1)}&\!=\!& 
u_n (u_{n+1} - u_{n-1} + v_{n+1} - v_n), \quad\quad\;\;
G_2^{(1)} \!=\! v_n (u_n - u_{n-1}), \\
G_1^{(2)}&\!=\!& 
u_n^2 (u_{n+1} \!- u_{n-1} \!+ v_{n+1} - v_n) \!+ u_n [(u_{n+1} \!+ v_{n+1})^2 
\!- (u_{n-1} + v_n)^2 \nonumber \\
&& + u_{n+1}(u_{n+2} + v_{n+2}) - u_{n-1} (u_{n-2} + v_{n-1})], \\
G_2^{(2)}&\!=\! & v_n^2 ( u_n - u_{n-1}) 
+ v_n (u_n^2 - u_{n-1} u_{n-2} - u_{n-1}^2 
+ u_n u_{n+1} \nonumber \\
&& - u_{n-1} v_{n-1} + u_n v_{n+1}).
\end{eqnarray}
Conserved densities and symmetries of other relativistic lattices are 
in \cite{baltzer,physletta}.
% \vfill
% \newpage
\section{Conclusions}
We presented three methods to test the integrability of differential
equations and difference-differential equations.
One of these methods is the Painlev\'e test, which is applicable to
polynomial systems of ODEs and PDEs.  

The two other methods are based on the principle of dilation invariance.
Thus far, they can only be applied to polynomial systems of evolution 
equations. As shown, it is easy to adapt these methods to the DDE case.  

Although restricted to polynomial equations, the techniques presented 
in this paper are algorithmic and have the advantage that they are 
fairly easy to implement in symbolic code. 

Applied to systems with parameters, the codes allow one to determine 
the conditions on the parameters so that the systems pass the Painlev\'e 
test, or have a sequence of conserved densities or higher-order symmetries.
Given a class of equations, the software can thus be used 
to pick out the candidates for complete integrability.

Currently, we are extending our algorithms to the symbolic computation of 
recursion operators of evolution equations.
In the future we will investigate generalizations of our methods to 
PDEs and DDEs in multiple space dimensions. The potential use 
of Lie-point symmetries other than dilation (scaling) symmetries 
will also be studied.

\section*{Acknowledgements}

We acknowledge helpful discussions with Profs. B. Herbst, 
M. Kruskal. S. Mikhailov, C. Nucci, J. Sanders, E. Van Vleck, F. Verheest, 
P. Winternitz, and T. Wolf. We also thank C. Elmer and G. Erdmann for 
help with parts of this project. 
% \vfill
% \newpage

\end{document}